I.I. Rakov[1], S.M. Pridvorova[2], G.A. Shafeev[1,3]

# Generation of nanoparticles of phtalocyanines by laser fragmentation and their interaction with gold nanoparticles

[1] A.M. Prokhorov General Physics Institute of the Russian Academy of Sciences, 38, Vavilov street, 119991 Moscow, Russian Federation

[2] A.N. Bach Institute of Biochemistry, Federal Research Centre «Fundamentals of Biotechnology» of the Russian Academy of Sciences, 33, Leninsky prospect, 119071, Moscow, Russian Federation

[3] National Research Nuclear University MEPhI (Moscow Engineering Physics Institute), 31, Kashirskoye highway, 115409, Moscow, Russian Federation

**Abstract**

Optical properties and morphology of laser generated Aluminum and Copper phthalocyanine nanoparticles (nAlPc and nCuPc) in water are experimentally studied. Near infrared laser source of nanosecond pulse duration was used for fragmentation of Pc micro-powder suspended in $H_2O$. Extinction spectra in the visible and near IR range of NPs colloidal solutions in MQ water were acquired by means of optical spectroscopy. The optical density of both nCuPc and nAlPc increases with laser fragmentation time. Transmission electron microscopy was used for characterization of nanoparticle morphology and size analysis. It is found that nCuPc are made of short (100 nm) rectangular bars interconnected at various angles with other bars. Similar experiments were carried out for a colloidal solution, which is a mixture of Au and AlPc nanoparticles. It turned out that Au NPs in presence of nAlPc form large agglomerates of Au.

**Introduction**

Major of phthalocyanines produced ~ 90% in the form of phthalocyanine complexes with transition metals is used as pigments. Sustained, intense and uniform color combined with the synthetic availability of these flat aromatic π-systems determines their widespread use in various industries (application in recording devices (CD / R), in liquid crystal displays, photoconductors in laser printers, and also as radiation absorbers and *p*-conductors in organic cells in solar cells.

In addition, they are used in the diagnosis of cancer (photosensitizers), photothermotherapy and subsequent photodynamic therapy, laser-induced fluorescence diagnostics (PD), etc [1-3]. It is required to synthesize phthalocyanines (Pc) of various metals and a specific morphology for each of these tasks. For these purposes Pc micro- and nanoparticles are mainly used in the form of powder or Pc colloidal solutions. Controlling the size of Pc nanoparticles is highly demanded for numerous applications. One of the possibilities to establish such control is laser-induced fragmentation of phthalocyanines suspensions. The process of laser-induced fragmentation of nanoparticles (NPs) is well known. This process has been thoroughly explored for metallic NPs in liquids. This concerns the interaction of individual NPs with laser beam in the liquid. As the result of this interaction, average size of NPs decreases with laser exposure time of the solution.

Under different experimental conditions this phenomenon may be caused by different processes, such as hydrodynamic instabilities or Coulomb explosion [4 - 7]. The nanoparticle fragmentation process was shown to depend on the initial particles size. Particularly, relatively small particles (less than 200-300 nm in diameter) are fragmented via separation of the finer fragments (of about 10 nm in diameter) in form of nanoparticles [8]. On the contrary, larger micrometer-sized particles under the influence of laser radiation are divided in equal parts [9]. It is pertinent to note that fragmented NPs have usually spherical shape, since the interaction of NPs with laser radiation proceeds through their melting.

Thus, this work is devoted to AlPc and CuPc micro-powder fragmentation process research. The nPc morphology and size changing by laser radiation discover new abilities for application in different areas.

## 1. Laser fragmentation of phtalocyanines

The technique of laser fragmentation in liquid was used to generate phthalocyanines (Pc) nanoparticles (NPs). In particular, the fragmentation scheme with different exposure times of the initial suspension of AlPc and CuPc NPs was realized. Ytterbium fiber laser with pulse duration of 100 ns, repetition rate of 20 kHz and pulse energy of 1 mJ at 1060−1070 nm was used as radiation source. Focusing was carried out by F-Theta objective with a focal length of 207 mm. The laser fluence inside the liquid was about 13-15 J/cm$^2$. This is much higher than the fluence used in previous work [1]. The laser beam was scanned inside the colloidal solution along the circular trajectory at a speed of 100 mm/s using a galvo-mirror system. Scanning laser beam induces circular convective flows in the liquid improving thus the agitation of the suspension [10]. Ultrapure Milli-Q (MQ) water was used as a working liquid. Extinction spectra of colloidal solutions were acquired using an Ocean Optics UV−Vis fiber spectrometer in the range of 200− 900 nm. The morphology of AlPc and CuPc NPs was analyzed by means of Transmission Electron Microscopy (TEM). Experimental setup is shown in Fig.1.

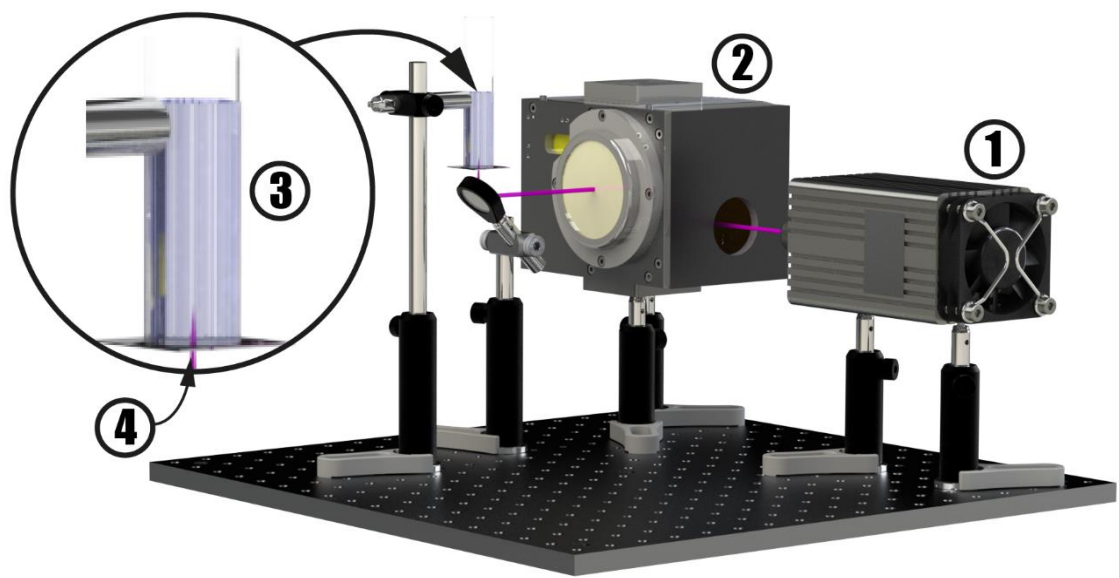

Fig. 1. Experimental setup for laser fragmentation of phthalocyanines (Pc); 1 – laser, 2 – galvo-optic system with F-Theta lens; 3 – cooled cuvette with Pc colloidal solution; 4 – scanning laser beam.

A series of experiments on mixing colloidal solutions of Au and AlPc nanoparticles was also conducted. In the first case, a mixture of nanoparticles colloidal solutions was irradiated for 20 minutes with similar laser parameters. In the second case, direct and reverse mixing of Au and AlPc colloidal solutions occurred directly in the spectrometer cuvette. In the case of direct mixing, phthalocyanine nanoparticles were added to the colloidal solution of gold nanoparticles in 25 µl increments of the dispenser. In the case of reverse mixing, gold nanoparticles were added to dilute AlPc NPs colloidal solution in 100 µl increments. Sartorius (Biohit) Proline 25-50 µl dispenser was used for this purpose.

Laser exposure of Pc micro-powder (either Al Pc or Cu Pc) is not accompanied by laser breakdown of the solution. Bright flashes occur from time to time and are probably due to the exposure of relatively large pieces of Pc. This is explained by matching laser wavelength into window of relative transparency of studied Al Pc suspension (Fig. 2).

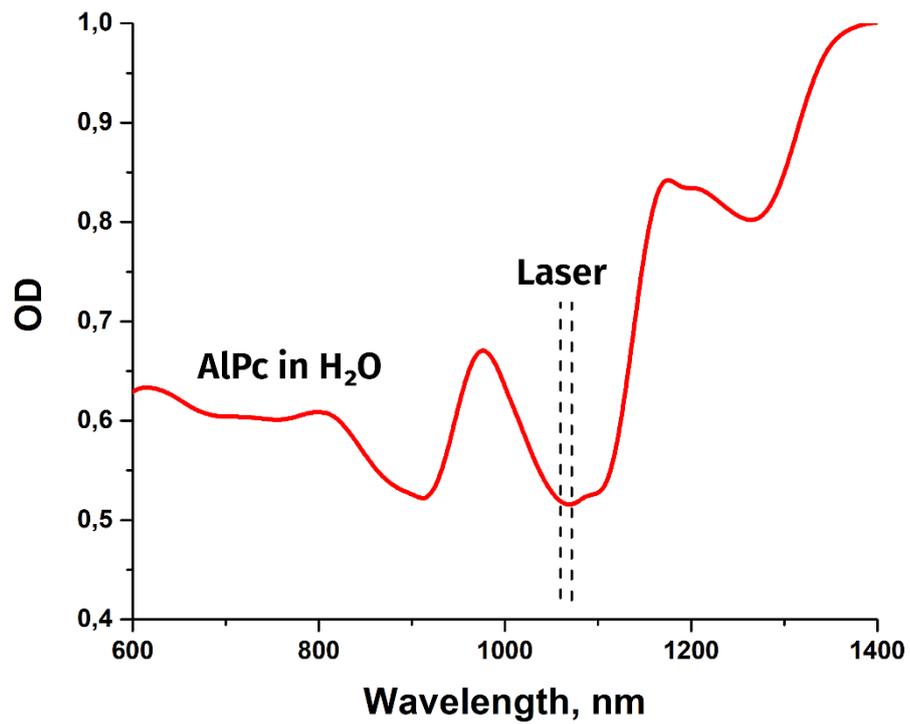

Fig. 2. Extinction spectrum of Al Pc NPs with respect to emission line of Ytterbium fiber laser.

Extinction spectra of both Al and Cu Pc at various stages of fragmentation are shown in Fig. 3.

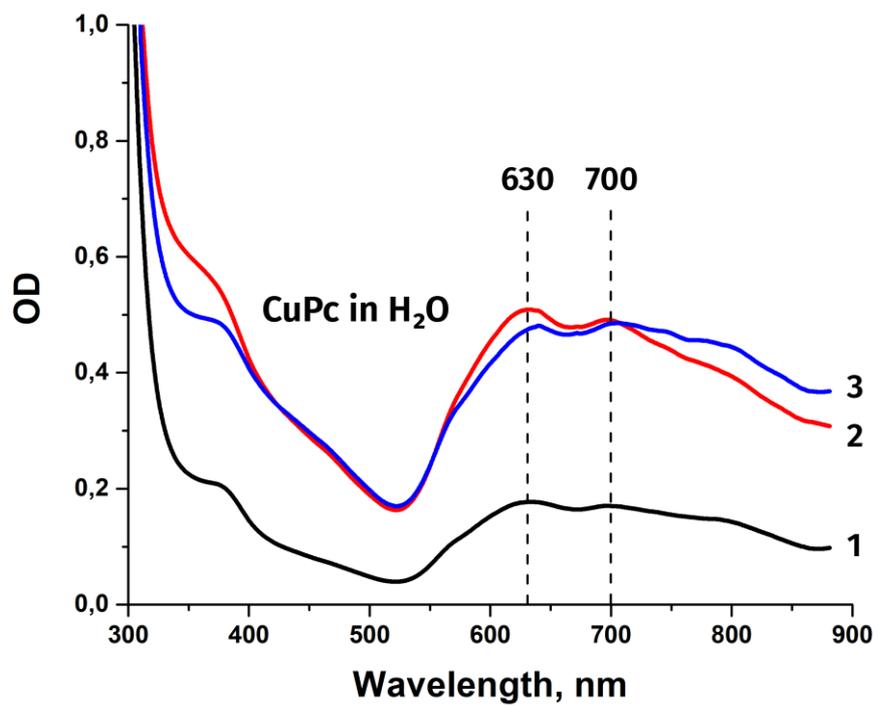

a

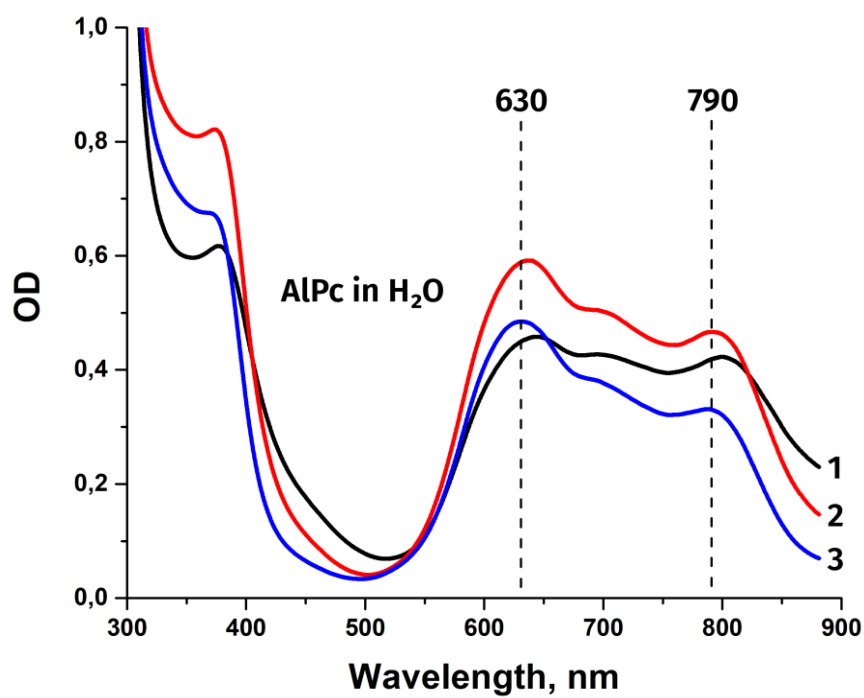

b

Fig. 3. Extinction spectra of laser-fragmented Cu Pc (a) and Al Pc (b) in water as the function of laser fragmentation time. 5 min of fragmentation (1), 10 (2) and 20 min (3).

One can see that the extinction spectrum of Cu Pc remains qualitatively the same with the increase of time of laser fragmentation. Optical density (OD) increases with time due to increase of concentration of Cu Pc nanoparticles. On the contrary, the ratio of peak intensities at 630 and 790 nm of Al Pc spectrum changes with fragmentation time. This indicates possible modification of molecular structure of Al Pc under laser radiation.

TEM images of Cu Pc NPs generated by laser fragmentation of initial micro-powder of Cu Pc are presented in Fig. 4.

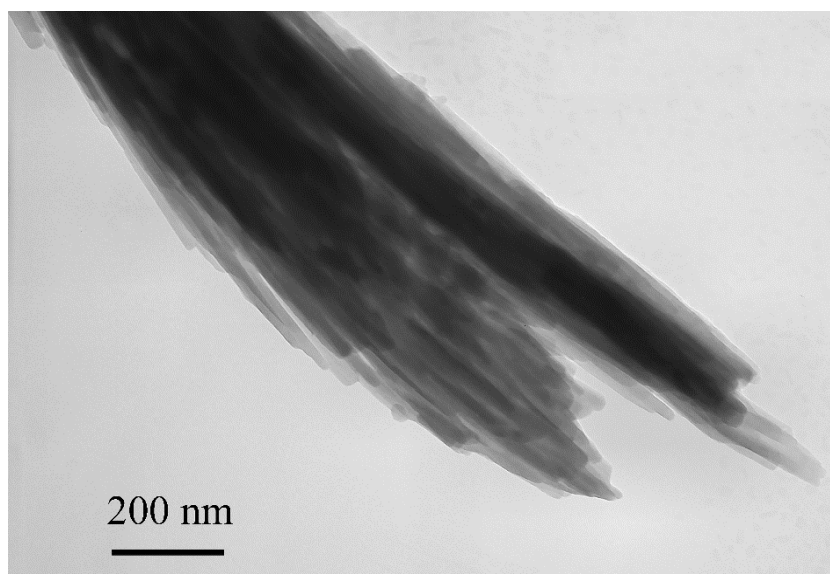

a

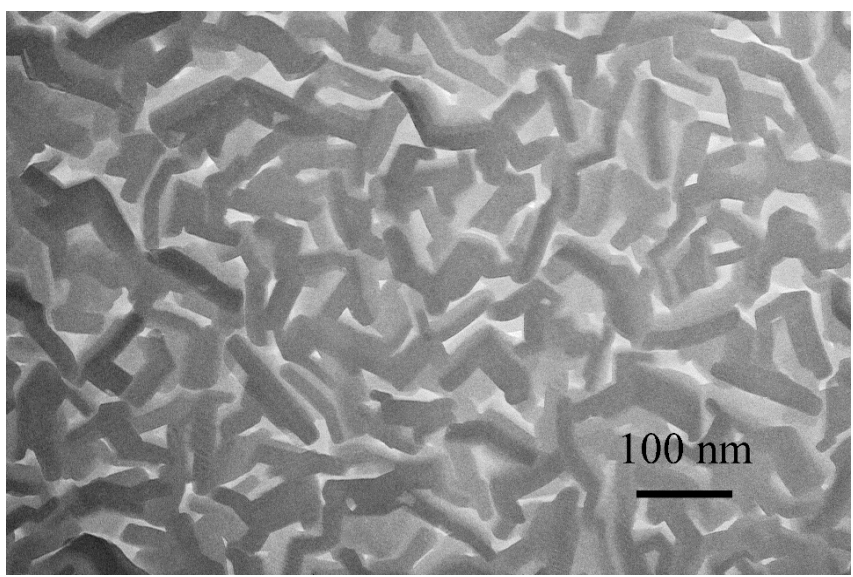

b

Fig. 4. NPs of Cu Pc at various stages of laser fragmentation. Initial stage (a) and after 10 min of fragmentation (b).

One can see that NPs have very peculiar shape and are not rounded as it is usually observed after laser ablation and fragmentation in liquids [5, 9]. They are made of straight bars of about 100 nm long and connected between each other at different angles. The lateral dimensions of these bars are about 30 – 50 nm. Higher magnification reveals the systems of parallel planes visible in some favorably oriented NPs (Fig. 5).

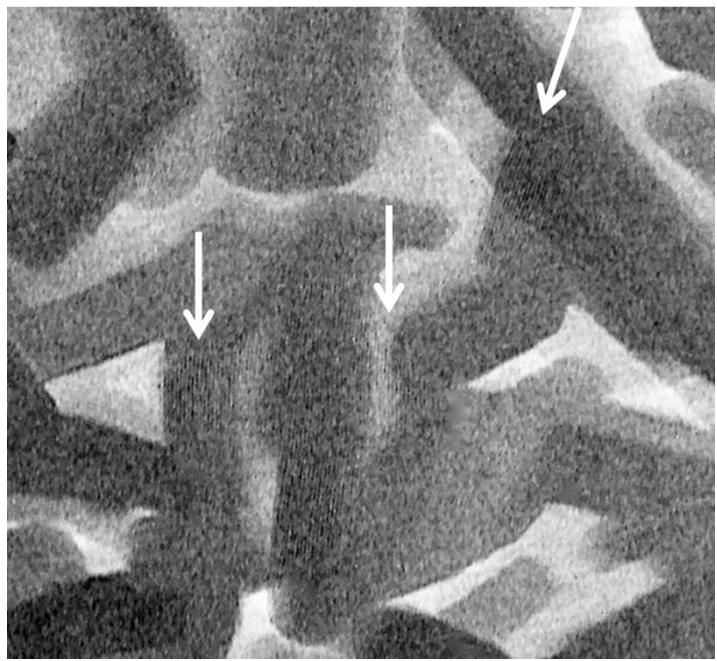

Fig. 5. Enlarged TEM view of Cu Pc nanoparticles. Arrows indicate the system of parallel planes with measured period of 0.9 nm.

At high temperature phthalocyanines are not molten but sublimed. Upon fast cooling they are crystallized mostly along directions of fastest growth. Molecules are added more rapidly to the edges and corners of growing crystal than to the centers of crystal faces, resulting in their tree-like forms. This kind of crystals is called skeletal ones.

## 2. Interaction of nanoparticles of Pcs with Au NPs

NPs of Al phthalocyanine generated by laser fragmentation in $H_2O$ were added by small portions to colloidal Au NPs generated by laser ablation of a bulk Au target in MQ water. The variations of the extinction spectrum are shown in Fig. 6. Addition of 100 μl of NPs of Al Pc generated by laser fragmentation leads to complete dumping of plasmonic band of Au NPs at 527 nm. The volume of Au NPs solution was about 2 $cm^3$, so such strong dumping of the plasmon resonance of Au NPs by μl additions of Al Pc NPs should be explained by incommensurable concentrations of both colloids.

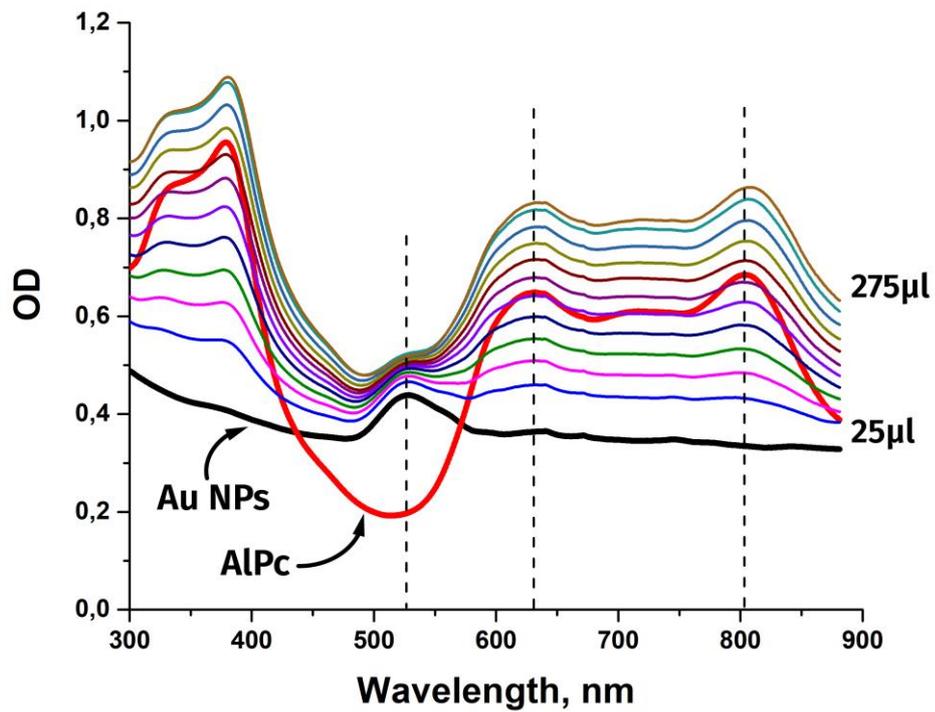

Fig. 6. Variation of extinction spectrum of Au NPs generated by laser ablation of gold target in water upon addition of colloidal solution of Al Pc NPs generated by laser fragmentation in $H_2O$.

In a wider spectral range the same evolution of the extinction spectrum looks as shown in Fig. 7.

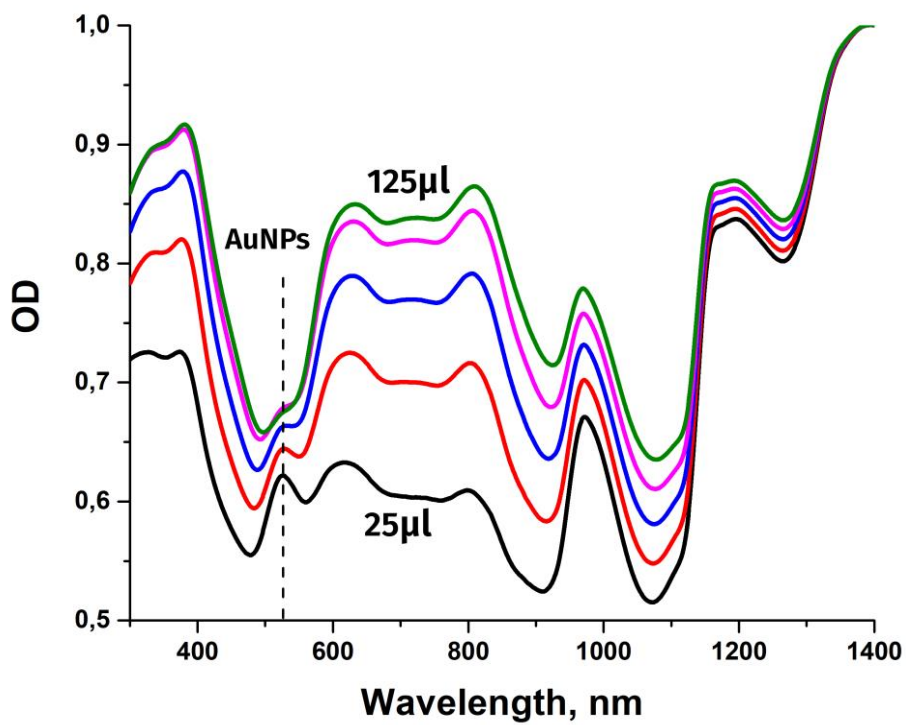

Fig. 7. Variation of extinction spectrum in the visible and near IR range of Au NPs generated by laser ablation of gold target in water upon addition of colloidal solution of Al Pc NPs generated by laser fragmentation in $H_2O$.

In the reverse mixing (Au colloid added to AlPc colloid) the sequence of extinction spectra is as follows in Fig. 8.

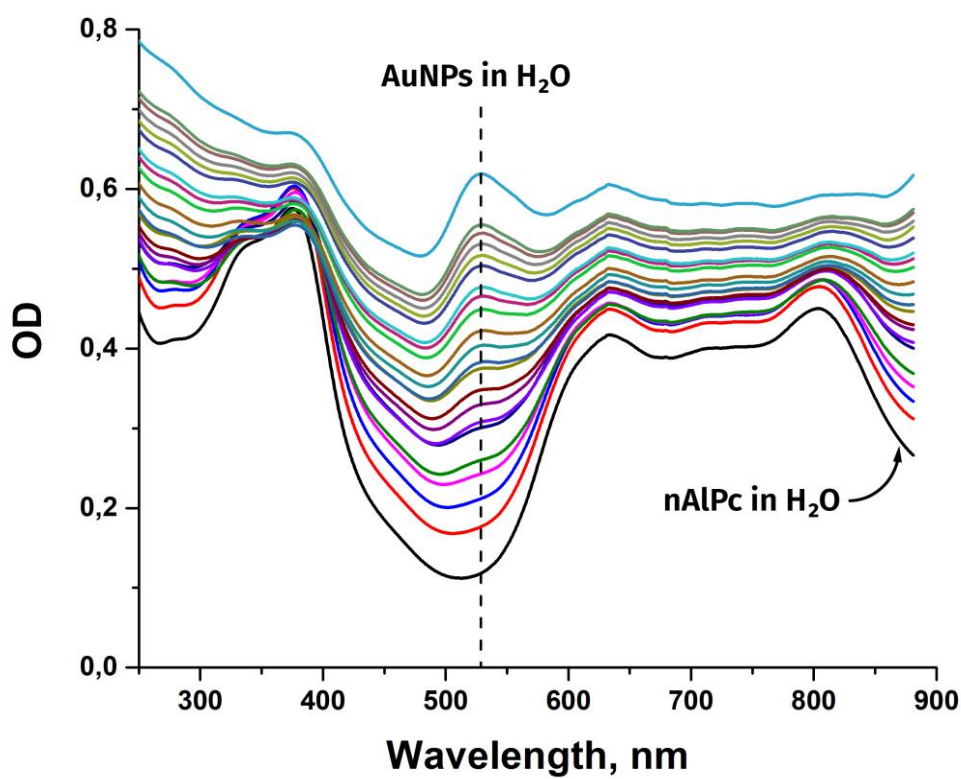

Fig. 8. Variation of extinction spectrum of Al Pc NPs in water upon gradual additions of 100 µl of colloidal solution of Au NPs in water. Upper spectrum corresponds to 3 ml of Au colloid added. Center of the plasmon band of Au NPs is indicated by vertical line.

Note that the optical density of the mixture starts increasing in the red region of spectrum (near 800 − 850 nm) upon addition of relatively large portions of Au NP solution. The increase of absorption in this region is characteristic of elongated Au NPs [11]. Therefore, one may conclude that nAlPc cause the elongation and aggregation of Au NPs.

**Conclusion**

Thus, laser-assisted fragmentation of phtalocyanines of both Cu and Al in water suspensions has been demonstrated. Optical absorption spectroscopy confirms the increase of concentration of

nPcs with laser exposure time. Extinction spectra of nPcs remain qualitatively the same throughout the laser exposure indicating negligible damage to the molecular structure of Pcs. Nanoparticles of phtalocyanines have the structure of skeletal crystals and are made of short bars around 100 nm long interconnected with other bars by shorter bars. Nanoparticles of phtalocyanines cause elongation and agglomeration of Au NPs generated by laser ablation in water.

## Acknowledgments

This work was performed within State Contract No. AAAA-A18-118021390190-1 and within the framework of National Research Nuclear University 'MEPhI' (Moscow Engineering Physics Institute) Academic Excellence Project (Contract No. 02.a03.21.0005) and supported in part by the Russian Foundation for Basic Research (Grant Nos 18-52-70012_e_Aziya_a, 18-32-01044_mol_a). The authors acknowledge the support from Presidium RAS Program No. 5: Photonic technologies in probing inhomogeneous media and biological objects.